\documentclass[conference]{IEEEtran}
\usepackage{color} 
\usepackage{fancyhdr}
\usepackage{cite}

\usepackage{graphicx}
\DeclareGraphicsExtensions{.eps}
\usepackage{setspace}
\providecommand{\PSforPDF}[1]{#1}
\usepackage[cmex10]{amsmath}
\usepackage{amsfonts,amssymb,amsthm}
\usepackage{times}
\PSforPDF{\usepackage{psfrag}}
\usepackage{algorithmic}
\usepackage{algorithm}
\usepackage{array}
\usepackage{balance}
\usepackage{multirow}
\usepackage{amssymb}
\usepackage{mdwmath}
\usepackage{url}
\usepackage{mdwtab}
\usepackage{eqparbox}
\usepackage{nicefrac}
\usepackage{soul,xcolor}
\setstcolor{magenta}
\usepackage{exscale,relsize}
\ifCLASSINFOpdf
\else
\fi
\begin{document}

\title{Energy saving market for mobile operators }

\author{\IEEEauthorblockN{M M Aftab Hossain}
\IEEEauthorblockA{School of Electrical Engineering\\
Aalto University, Finland\\
Email: mm.hossain@aalto.fi}
\and
\IEEEauthorblockN{Riku J\"antti}
\IEEEauthorblockA{School of Electrical Engineering\\
Aalto University, Finland\\
Email:riku.jantti@aalto.fi}
\and
\IEEEauthorblockN{Cicek Cavdar}
Wireless@KTH, \\
KTH Royal Institute of Technology, Sweden\\
Email:cavdar@kth.se}
\maketitle

\begin{abstract}
Ensuring seamless coverage accounts for the lion's share of the energy consumed in a mobile network. Overlapping coverage of three to five mobile network operators (MNOs) results in enormous amount of energy waste which is avoidable. The traffic demands of the mobile networks vary significantly throughout the day. As the offered load for all  networks are not same at a given time and the differences in energy consumption at different loads are significant, multi-MNO capacity/coverage sharing can dramatically reduce energy consumption of mobile networks and provide the MNOs a cost effective means to cope with the exponential growth of traffic. In this paper, we propose an energy saving  market for a multi-MNO network scenario. As the competing MNOs are not comfortable with information sharing, we propose a double auction clearinghouse market mechanism where MNOs sell and buy capacity in order to minimize energy consumption. In our setting, each MNO proposes its bids and asks simultaneously for buying and selling multi-unit capacities respectively to an independent auctioneer, i.e., clearinghouse and ends up either as a buyer or as a seller in each round.  We show that the mechanism allows the MNOs to save significant percentage of energy cost throughout a wide range of network load. Different than other energy saving features such as cell sleep or antenna muting which can not be enabled at heavy traffic load, dynamic capacity sharing allows MNOs to handle traffic bursts with energy saving opportunity.
\end{abstract}

\section{Introduction}
The current trend indicates that the global mobile data traffic will keep growing exponentially and will  increase $11$ times by $2018$ compared to  what it was in $2013$\cite{Cisco}.  The corresponding increase in energy consumption is untenable from business perspective as it is understood that the revenue from meeting this extra capacity demand will not increase significantly or may not increase at all. This revenue gap poses a huge challenge to the MNOs to come up with sustainable business models. Business viability as well as environmental awareness calls for system and solutions that can cater high capacity demand with manifold increase in energy efficiency.  As network load varies throughout the day (i.e., the daily maximum are  even $2-10$ times higher than the daily minimums) and the  variation of  load demands among the MNOs  serving the same geographical area is significant \cite{Teeraparpwong},   the MNOs can utilize this load behavior in their favor if they share capacity.  Multi-MNO capacity sharing can be a cost-effective and necessary means to restrain energy consumption to support exponential traffic growth. However, this necessitates redefining the relations among the regulator and competing MNOs.

Recently, energy efficiency in wireless network has garnered significant attention and  different solutions have been proposed. There are numerous studies that suggest dynamic cell range so that a portion of the network can be switched off when network load becomes low \cite{Yong,  Marsan3}. However, it is very difficult to maintain proper coverage while sending a set of  base stations (BSs)  offline or in sleep mode. 
 
Mobile data offloading is another viable option in order to cope with the increasing traffic demands and to refrain from additional capacity deployment. 
 Many studies \cite{Iosifidis, Dimatteo, Lin} demonstrated the benefit of Wi-Fi offloading. However, \cite{PalVtc} has shown that the total energy consumed by a network for data transmissions is in the single digit percentage and the rest is mainly consumed  by the BSs to ensure coverage. As a result, offloading to only  Wi-Fi does not improve the situation much unless all data can be offloaded and the BS switched off. Deployment of small cells is also not that effective for the same reason. Further complementing cost and energy saving mechanisms are required.

Different forms and modes of multi-MNO sharing have been discussed recently. The proposals range from sharing  few sub-carriers to sharing everything, e.g., infrastructure, spectrum, capacity. In \cite{Marsan1,  Cosimo, Niu}, cooperation among multi-MNOs has been suggested in order to accomplish greener operation.  In \cite{Marsan1}, it has been suggested that the MNOs can cooperate during off-peak hours and reduce energy cost by shutting down one MNO through offloading its users to the other MNO. In \cite{Cosimo}, under game-theoretic framework, formation of stable coalitions among MNOs has been proposed in order to maintain QoS and reduce energy consumption by sharing both users and BSs. In \cite{Niu}, different possibilities including cell switching off and inter-MNO cooperation has been discussed. However, the MNOs may consider their information sensitive and  not be willing to share  with their competitors. As a result, a collaborative scheme that requires sharing of information is difficult to be implemented in reality. 

Double auction (DA) is an appropriate mechanism for  a scenario in which the number  of both sellers and  buyers are more than one and none of them is willing to reveal information about demand and supply. Sellers compete with each other in order to attract buyers and the buyers compete among themselves and can offer bids for some or all the sellers. Normally, an independent auctioneer collects the bids and asks from the buyers and sellers respectively, selects the winning sellers and buyers, allocates the items  from the sellers to the buyers and determines prices from the buyers to the sellers~\cite{Iosifidis2}. DA has been proposed extensively  in wireless networks as the  mechanism to solve different problems, e.g., power and spectrum allocation, spectrum sharing for both  intra-MNO and inter-MNO setting \cite{Iosifidis},\cite{Wenchao,  Iosifidis3}. We adopt this mechanism  in order to design a multi-MNO energy saving  market. 

 In this paper, we investigate the energy saving potential by allowing the MNOs to trade capacity by establishing an energy market.  The motivation behind this is to utilize the existing capacity of the MNOs to the fullest in order to minimize energy consumption. Under this mechanism, the MNOs  utilize the free capacity of their competitors by traffic offloading  and  achieve mutual benefits as the cost function for accommodating excess traffic is convex and the utility function of the traffic offloading is concave~\cite{Iosifidis,Costas}. The networks are interference limited at busy hour and energy consumption increases almost exponentially when the network load approaches to the peak. However, the difference in energy consumption for serving an extra unit of traffic  at different loading condition is significant as shown in Figure $1$. Let the current offered load of MNO $A$ be $L_1$ and MNO $B$ be $L_2$. The bid and ask by the MNO $A$ for $x$ unit has been shown in the Figure $1$. Note that by offloading $x$ unit of load, energy savings achieved by MNO $A$ is larger than the energy cost conceded by MNO $B$ to serve that extra $x$ unit.
\begin{figure}[t]
\centering
        \includegraphics[ scale=.65]{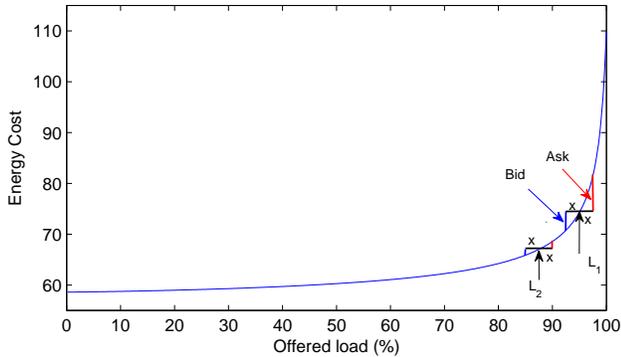}	
\caption{Energy cost for one MNO vs. offered traffic load, normalized to the feasible load for one BS.}
\label{fig:Costfunction}
\end{figure}

As a result, there is an economic viability to share capacity. This capacity sharing also allows MNOs to be less stringent with over dimensioning in order to handle the bursty  traffic demand during the busy hours. On the other hand, at low load, one or more MNOs can offload totally and go offline or to a deep sleep mode and save significant amount of energy (e.g., see Figure $1$).  As the MNOs are not willing to reveal their confidential information, we propose a clear-house DA as the market mechanism.  Numerical results suggest that the energy saving potential is  significant at low to medium load. Also, the amount of energy saved during busy hour is  considerable, especially, when one or two MNOs are highly loaded compared to other MNOs.

The rest of the paper is organized as follows: In Section \ref{sec:SM}, we provide the system model. 
 In Section \ref{sec:EM}, we present the energy saving market. In Section \ref{sec:NA}, we illustrate the numerical analysis.  Our paper closes with a conclusion and future work in Section \ref{sec:Conclusion}.

\section{ model and assumptions}
\label{sec:SM}
Consider the down-link of  $M$ MNOs whose serving areas are overlapping in the same geographical area and  each MNO consist of a set $\mathcal{K}= \{1,2, ... K\}$ of cells. A BS serves only one cell, hence BS and cell is used interchangeably hereafter. Each BS is assumed to have its own power amplifier (PA).  We consider that each BS is similarly loaded and hence  the central cell represents the network of an MNO. However, while calculating interference for the central cell, interference from all the neighboring cells of its own MNO has been accounted for. 
  We model the BS as a single server processor sharing (PS) queue, i.e., $M/G/1$-PS queue that serves the users in one-by-one fashion. For simplicity, we divide space into discrete locations. Let $u\in\mathcal{C}_k$ denote the location of any user in  the finite set of locations $\mathcal{C}_{k}$  in the cell $k \in \mathcal{K}$. Flows are generated in the BS with rate $\lambda_u$  packets per second and packet size $S_u$ for the location $u$.  The transmission rate for the user at location $u$, $r_u$ is a function of the BS transmit power $p$ and the interference received, $I_u$. The service time for a packet at location $u$  is $x_u= S_u r_u^{-1}$. Let $P_{PA}$ denote the total power consumed by the PA of the BS for the fixed downlink transmit power $p$. Note that we do not consider any downlink power control which is inline with current technology, LTE. The consumed power other than the PA consumption during the transmission  and during the idle state is denoted by  $P_c$ and  $P_{idle}$ respectively. 

The energy consumed to transmit a  packet at location $u$ is 
\begin{equation}
e_{tx}(S_u)=(P_{PA}+P_c) x_{u}
\nonumber\\
\end{equation}

The total energy consumption  for serving offered load, $L = \sum_u S_u\lambda_u$ including the energy consumed during idle time 
\begin{eqnarray} 
e_{tot}(L)&=&\sum_{u\in\mathcal{C}}{\lambda_{u}e_{tx}(S_u)}+\Big(1-\sum_{u\in\mathcal{C}}\lambda_{u}x_{u}\Big) P_{idle}\nonumber \\
&=& \sum_{u\in\mathcal{C}}\lambda_{u} S_{u} \frac{P_{PA}+P_c-P_{idle}}{r_u}+P_{idle} 
\label{eq:Energy}
\end{eqnarray}
where  $\sum_{u\in\mathcal{C}} \frac{\lambda_{u} S_{u}}{r_{u}} < 1$ is the queue stability constraint. 
 Note that  the fraction of time the server is busy, i.e.,  $\sum_{u\in\mathcal{C}} \frac{\lambda_{u} S_{u}}{r_{u}}$, is the  activity of  a BS and is coupled among BSs due to the dependence of achieved rate on received interference from neighboring cells. The activity is determined through an iterative process. For a given offered load, in order to calculate the activity of a cell,  the activity of all other interfering BSs are initialized as $1$, i.e., BSs are considered always active. The iterations are carried out until the activity of all  cells converges. 
 
\subsection{Power amplifier}
 The traditional power amplifier (TPA) is characterized by high efficiency only close to its compression region \cite{Hossain}. The required input power for mean transmit power level equal to $p\leq p_{max}$ is 
\begin{equation}
P_{PA}(p) = \frac{1}{\eta_{max}} \sqrt{p\cdot p_{max,PA}}
\label{eq:TPA}
\end{equation} 
 $\eta_{max}$ denotes the maximum PA efficiency while delivering maximum PA output power, $p_{max,PA}$. Note that for modern technologies, e.g., CDMA,OFDM, maximum average  transmit power of a BS ( i.e., $p_{max}$ ) needs to be around $8$ dB less  than $p_{max,PA}$ due to high peak to average power ratio (PAPR).

\subsection{Aggregate interference model and transmission rate}
We consider log-normal shadow fading and  we employ the Fenton-Wilkinson approximation method as in~\cite{OurJournal} in order to model the aggregate interference level in the slow fading environment. In the downlink, the locations of the interfering BS are fixed. The mean and variance of the aggregate interference received by a mobile terminal (MT) at a location can be found at~\cite[Eq. (29), Eq. (30)]{OurJournal}. For modeling the transmission rate we consider that the users can adapt their  modulation and coding scheme so that they can reach the Shannon capacity bound.

%
\section{MNO Energy market} 
\label{sec:EM}
This  energy market models the trading of network capacity among the cellular MNOs in order to minimize the network energy consumption.  One particular key feature of our energy market is that each MNO submits both bids, i.e., offers to buy and asks, i.e., offers to sell simultaneously to a DA clearinghouse with the view to maximize their profit. The MNOs participate in the auction repeatedly and in a single round an MNO is allowed  either to buy or sell.  The MNOs are expected to submit offers for  multi-unit capacities and corresponding differentiated prices. 
 In each round, the MNOs revise their  bids  and asks based on their current offered loads.  There are some parameter values for each MNO, e.g., the unit of load it wants to trade and the anticipated trade price.  The values of these  parameters are private to each MNO and  hidden from each other.   The clearinghouse collects the bids and asks from each MNO and  matches them following its criterion to minimize total energy consumption, determines the winning buyer and seller MNOs, trade price and quantity.  

In order to generate  the bid to offload a unit of offered load,  an MNO determines how much energy it can save by offloading that unit. Similarly, to generate the asks it calculates the energy cost it suffers while accommodating that additional unit. The clearinghouse use  Preston McAfees DA (PMD) protocol to determine the trade price, $p_{trade}$ \cite{PMD} and allocation process. In order to use this protocol we consider each unit of load as a single item and each ask or bid from the same MNO are considered as an independent bid or ask. The reason behind using PMD protocol is that  it has the properties of being  i) dominant-strategy incentive compatible, i.e., truthful bidding is the best strategy for the bidders, ii) budget balanced: auctioneer ends up with non-negative payments  and iii)individual rational: bidders do not get worse by participating. However, PMD cannot always ensure maximum social welfare, i.e., efficiency as will be described later. Note that no double auction mechanism can ensure all these properties and other DA protocols yield similar results in this study.   In PMD protocol, bids ($\textbf{b}$) and asks ($\textbf{a}$) are arranged in descending and ascending order respectively. Then lowest bid, $b_{j}$ is identified such that  $b_{j} \ge a_{j}$ and $b_{j+1} < a_{j+1}$. The trade  price that clears the market is determined as
\begin{equation}
p_{trade} = \frac{1}{2}( b_{j+1}+a_{j+1})
\end{equation} 
If $b_{j}\ge p_{trade} \ge a_{j}$, the traded quantity is $j$ and if  $b_{j} \le p_{trade}$ or $p_{trade} \le a_{j}$, the traded quantity is $j-1$. The winning buyer for the last unit pays $b_{j}$ and the winning seller for the last unit receives $a_{j}$.

\subsection{Bid generation}
For traditional goods, the marginal utility diminishes with the increase of quantity. By offloading data an MNO saves a part of the dynamic energy, $P_{PA}$,  required to support that amount of data. This potential of energy saving decreases with the increase of offloaded amount. However, if the MNO can offload totally, it can save the static part of the energy (i.e. $P_{idle}$) by completely shutting down or going to deep sleep mode.
This fact leads to a particular bidding from the MNOs which keep decreasing with each extra unit of capacity followed by a drastic increase of total offer when it bids for total offloading. The bid for $m$-th unit of  excess capacity offered by $i$-th MNO can be given by 

\[b_{i,m} = \left\{
  \begin{array}{lr}
    e_{tot}(L_{i,m}) -e_{tot}(L_{i,m-1})-e_{tr} : \Delta l < L_{i,m}\\
     \end{array}
\right.
\]

where $L_{i,m-1} =L_{i,m}-\Delta l$, $e_{tot}(L_{i,m})$ is the energy consumption of the network with load  $L_{i,m}$ and $e_{tot}(L_{i,m-1})$  is the energy consumption of the network when it offloads unit offered load, $\Delta l$. $e_{tr}$ gives the energy consumed while offloading the traffic to the  target MNO. The energy for zero load is given by $P_{sleep}$ and for total offloading case, the bid for total offloading equals the total energy required to serve current load. We consider $P_{sleep}=0$ in this study.

\subsection{Ask generation}
Similar to bidding process, in order to generate the asks, the MNOs calculate how much excess energy cost is suffered  by conceding the load from the other MNOs. Note that the excess energy required to support the excess traffic depends on the current load the network is serving.
Also, as shown in Figure $1$, asks for conceding a unit of excess traffic keep increasing. We denote the ask by MNO $i$ for the $n$-th unit of offered load as $a_{in}$  such that 
\begin{equation}
a_{in} = e_{tot}(L_{i,n+1}) -e_{tot}(L_{i,n})+e_{tr}
\end{equation}
where $L_{i,n+1} =L_{i,n}+\Delta l$, $e_{tot}(L_{i,n+1})$ is the energy consumption of the network with offered load  $L_{i,n+1}$ and $e_{tot}(L_{i,n})$  is the energy consumption of the network with offered load $L_{i,n}$. Note that  $a_{i,n} < a_{i,n+1}$.

\subsection{Clearinghouse mechanism}
\label{Clhouse}
The target of the auctioneer is to achieve the socially optimal allocation, i.e., minimization of  total energy consumption in the  service area.  In order to find the optimal  unit of offloaded traffic, $P$ and received excess traffic, $Q$ by the buying and selling MNOs, respectively,  clearinghouse needs to solve  the following social welfare maximization problem: 

 \begin{subequations} 
\label{eq:P1}
\renewcommand\theequation{\theparentequation\roman{equation}}
\begin{align} 
   {\mathop {{\text{Maximize}}:}_{P, Q}} &\;\;
   {\sum_{i}  
      B_{im}  p_{im} -\sum_i 
      A_{in}  q_{in}} \label{eq:P11}\\
		{{\text{Subject to:}}} &\;\; {\sum_{m } { p_{im}}+\sum_{n }  q_{in} \le 1 }, \forall i  \label{eq:P12}\\
		 &\;\;  {\sum_{i }\sum_{m } {mp_{im}}=\sum_{i }\sum_{n}n q_{in}} \label{eq:P13}.
		\end{align}
		\end{subequations}
where $A_{in} = \sum_n{a_{in}}$ is the total  ask for $n$ units of capacity submitted by MNO $i$, $B_{im} = \sum_m{b_{i,m}} $ is the  bid  for  offloading $m$ units of traffic by MNO $i$. This objective function corresponds to total energy savings. $p_{i,m} \in \{0,1\}$  and $q_{i,n} \in \{0,1\}$ are the decision variables where $p_{im}$ denotes whether ($p_{im}= 1$) or not ($p_{im}= 0$) $m$ unit of capacity has been allocated for MNO $i$.   Constraint (\ref{eq:P12}) ensures that an MNO ends up only as a buyer or seller and constraint (\ref{eq:P13}) makes sure that total amount sold are equal to total amount bought. As the number of MNOs covering same geographical area is small (i.e. $2-5$) in reality, solving this problem by brute force is not computationally demanding.
	
There are  possible  $\sum_{i=1}^{n-1} C_{n,i} = 2^n-2$  combination of MNO that can offload totally. For each combination first we i) update the asks by excluding the own asks of the offloading MNOs ii) arrange the bids in descending and the asks in ascending order iii) check if there are enough valid (i.e., bid is higher than the corresponding ask) asks.  In the second step, PMD protocol is used to complete the allocation process and determining the trade price, $p_{trade}$ from the rest of the bids and asks.  Finally, we combine the allocations and choose the combination that maximizes (\ref{eq:P11}). We update the allocation vectors $P$ and $Q$. However, if no MNO can offload totally, we simply use PMD protocol to make the allocation and determine trade price as described earlier.  We provide the pseudocode of the  algorithm to find the allocation vectors, $P, Q$ and trade price, $p_{trade}$.

\begin{algorithm} 
\label{algo:TOA}
\caption{Clearinghouse Algorithm}
\begin{algorithmic}[1]
\STATE Set of MNOs, $S_O$
\STATE Sets of offloading MNOs $S_B \leftarrow \{S_1, S_2, S_3 . . . S_K\}$
\STATE  Sets of capacity selling MNOs $S_{Ak} = S_O\setminus S_k, k=1,..., K$,  $S_A \leftarrow \{S_{A1},S_{A2},...,S_{AK}\}$
\STATE $\textnormal{Sets of  bids for the subsets of} \, S_B$ from $b_{i,m}$ 
\STATE $B\leftarrow \{b_1, b_2, . . . , b_K\}$, $b_k$ = the bids for the MNOs, $S_k$
\STATE $\textnormal{Sets of  asks for the subsets of} \, S_A$ from $a_{i,n}$ 
\STATE $A \leftarrow \{a_1, a_2 . . . a_K\}$ $a_k$ = valid asks for the MNOs, $S_k$
\STATE $S_I 	\subset	S_B$ where $1$-norm $ || b_i||_1 > ||{a_i}||_1$ and cardinality $|a_i| \ge |b_i|$
\IF   {$ \exists {S_I}$}
\STATE  Welfare $\leftarrow 0$
\FORALL  { ${S_I}$ }
\STATE  {Use PMD  for rest of the bids, $b^{'}_{i,m}$ and asks, $a^{'}_{in}$}
\STATE  Allocation vectors  $P$, $Q$ for winner traders
\STATE  Temp $\leftarrow$ Evaluate (\ref{eq:P11}) given $P$ and $Q$
\IF     {Temp $>$ Welfare}
\STATE Welfare $\leftarrow$ Temp, Print $P$, $Q$
\ENDIF
\ENDFOR
\ELSE 
\STATE  Use PMD protocol for all bids $b_{i,m}$ and all asks $a_{i,n}$
\STATE Allocation vector  $P$, and $Q$ for winner traders
\ENDIF
\end{algorithmic} 
\end{algorithm}

An MNO never shows up as both winning bidder and winning seller in the same round. Due to the exponential increase in energy cost, as shown in Figure $1$,  at any offered load, the ask for the first unit of an MNO is higher than it's bid for any unit except in the case of total offloading when the load becomes $0$. Therefore total offloading has been taken care of in the first part of the allocation process.
\section{numerical analysis}
\label{sec:NA}
We consider the network of each MNO consisting of $19$ regular hexagonal cells where wrap around technique has been employed to avoid border effects.  We also assume that the BSs of different MNOs are co-located and traffic distribution is homogenous in the coverage area.  Hence, the energy consumption in the  central cell represents the average energy consumption of the network.  A grid of $64$ points emulating the possible locations of users is generated inside each cell. As the energy required for processing packets at the BS is negligible, we consider the power consumed during idle mode  and the static energy consumption  during transmission as equal, i.e., $P_c=P_{idle}$. For TPA  we use $P_{idle}= 58.6$W  as derived in \cite {wcnc} based on the values given in \cite{EarthD2.3}.  The parameter settings for a network are summarized in the Table~\ref{RP}. 

\begin{table}[!ht]
\renewcommand{\arraystretch}{1.3}
\caption{simulation Parameters}
\label{RP} \centering
\begin{tabular}{|l|l|} \hline
\multicolumn {2} {|c|}{\textbf {Reference parameters }} \\
 \hline
  \textit{Parameter} & \textit{Value} \\
 \hline
 Number of cells & $19$ \\
 \hline
Grid size inside each cell & $64$ points \\
 \hline
Cell radius & $1$ km \\
\hline
PA maximum output power, $p_{max,PA}$ & $53$ dBm \\
\hline
Maximum average BS transmit power, $p$  & $46$ dBm\\
 \hline
Maximum PA efficiency at $p_{max,PA}$ & $80$\%\\
 \hline
Path loss exponent & $3.6$\\
 \hline
Shadow fading standard deviation  & $5.5$ dB\\
 \hline
Bandwidth  & $20$ MHz\\
 \hline
Noise level  & -$106$ dBm\\
 \hline
Target outage  & $10$\%\\
\hline
\end{tabular}
\end {table}

With the parameters provided in  the Table~\ref{RP}, we find  the cost function for generating each extra unit of traffic by a network in Figure  \ref{fig:Costfunction}. From this figure one can see that network energy cost remains relatively low even up to $80\%$ load. As a result a network can accommodate excess traffic from other MNOs without suffering much energy cost while operating on a wide range of its load. On the other hand, the energy saving potential is very high when the network can offload a fraction of its load when it is serving peak load.
\subsection{Energy saving potential at low to medium load}
\begin{figure}[t]
\centering
        \includegraphics[ scale=.6]{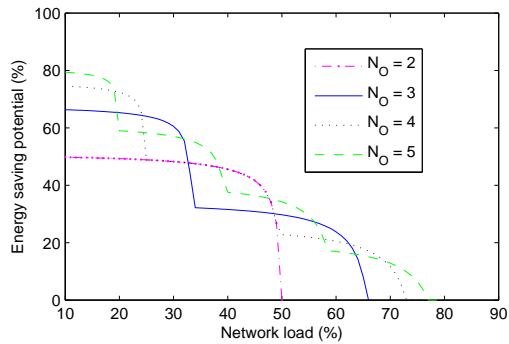}	
\caption{Energy saving potential by totall offloading of MNOs}
\label{fig:TotalOffload}
\end{figure}

In order to get some first insight into the energy saving potential by capacity sharing, we consider the loads of each MNO to be the same. From the Figure \ref{fig:TotalOffload}, one can see that the energy saving potential is significant if the networks are not highly loaded. When the network loads are  low to medium, energy is saved from total offload of one or more MNOs. When the network loads are very  high it does not give opportunity to offload any MNO totally and hence the energy saving potential ceases to exist.
Another important observation is that the energy saving potential increases with the number of MNOs involved in capacity sharing. With only two MNOs, maximum energy saving potential is around $50$\% whereas with five MNOs it jumps to $80$\%. The reason behind this is that if number of MNOs involved is bigger, more MNOs can be totally offloaded at low load. Also, for bigger number of MNOs involved, e.g., $5$,  it is possible to save energy by total offloading even at high loading condition of the networks. 
The drastic fall of energy saving potential happens when an MNO  need to be activated from sleep mode due to the increase in total load. 

\subsection{Energy saving potential at high load}
When the networks are highly loaded,  it is not possible to completely offload any  MNO. In such case, the capacity trading only helps the comparatively highly loaded MNOs to offload a fraction of their load and save from the dynamic part of the energy consumption, i.e., $P_{PA}$. In Table~\ref{Op5}, we present the energy saving potential when five MNOs share loads with some combinations of high loads. From the first two rows, one can observe that the percentage of energy saving potential becomes low with the combination of loads when no MNO can offload totally.  The next two rows show  higher saving percent as it is possible to offload at least one MNO at these moderate high loads. Note that although the percentage saving in energy consumption is comparatively low in case of very high network load, the actual amount of energy savings is considerable as energy consumption while serving high load is much higher. The last  row shows high energy saving potential if the network load of one or two MNOs are extremely high and others are moderate.  Although the networks do not operate with this extreme high loads, the main motivation of presenting these results is that with dynamic capacity sharing, the network dimensioning can be less stringent to cater the bursty highest load demand. MNOs dimension their network based on their busy hour traffic demand. As a result they end up with huge excess capacity for most of the time of the day. If multi-MNO capacity sharing is allowed, MNOs can reduce their CAPEX by avoiding overdimensiong and still maintain QoS by taking advantage of the spatial and temporal variation of load demands among themselves. 

\balance
From the numerical analysis, we also observe that if $e_{tr}$ is ignored, the networks become equally loaded after the trading of loads. This implies that balancing the loads among the MNOs is the appropriate strategy to minimize energy when considering equally loaded networks. In  \cite{Conte}, it has been shown that a BS can go to sleep by offloading existing users to other cells in around $30$ seconds. However, the frequency of auction round will also depend on the  time required for some meaningful change in the network load dynamics.

\begin{table}[!t]
\renewcommand{\arraystretch}{1.3}
\caption{Energy saving at high load 2}
\label{Op5} \centering
\begin{tabular}{|c|c|c|c|c|c|}
\hline
\multicolumn{5}{|c|}{\textbf{Operators load (\%)} } & {\textbf{Saving(\%)}}\\
 \cline{1-6}
\hline
90 &90 & 85 &65  & 60  & 2\\
\hline
95 &90 & 80 &70  & 70  &3\\
 \hline
 \hline
80 &80 & 80 &70  & 50  & 9\\
 \hline
85 &70 & 70 &60  & 50  & 16\\
 \hline
\hline
99 & 99 & 99 & 70 & 50 & 18\\

 \hline
\end{tabular}
\end {table}

\section{Conclusion}
\label{sec:Conclusion}
Efficient utilization of existing resources is a prerequisite in order to contain the operation cost of the MNOs, especially  when they are going to handle data avalanche with almost zero marginal revenue. By adopting appropriate mechanism and policy, e.g., national roaming, multi-MNO capacity sharing, it is possible to reduce network cost and inflict less harm to the environment by reducing ${\text{CO}_2}$ emission. In this paper, we have shown that double auction is a suitable mechanism for multi-MNO capacity sharing as it does not require sharing of private information. The energy saving potential has been found to be very high at low to medium load where one or more MNOs can be offloaded totally. Also, at high loads, considerable amount of energy can be saved if there is good gap among the MNOs loads.  It was also observed that for symmetric networks, energy efficient allocation balances load among the MNOs. In this work, we focused on the energy cost and did not consider quality of service (QoS) when generating bids and asks. However, QoS can be easily incorporated in the process.  Our assumption of co-located BSs is also the worst case scenario as connecting to a closer BS requires less energy to transfer same amount of data.  Also, we did not consider spatial variation of network load. For future work spatial variation of network load can be considered which will allow cell level offloading.   
\label{sec:CC}

\end{document}